\documentclass[9pt,twocolumn,twoside]{opticajnl}
\journal{opticajournal} 

\setboolean{shortarticle}{true}
\title{%Phase-Shift Induced Control of Supercontinuum Generation in Strongly Coupled Waveguides*** 
On-Chip Phase-Shift Induced Control of Supercontinuum Generation in a Dual-Core Si$\mathbf{_{3}}$N$\mathbf{_{4}}$ Waveguide}

\author[1,*]{L. Xia}
\author[1]{P.J.M. van der Slot}
\author[2]{M. Timmerkamp}
\author[1]{H.M.J. Bastiaens}
\author[2]{C. Fallnich}
\author[1,2]{K.-J. Boller}

\affil[1]{Laser Physics and Nonlinear Optics Group, Applied Nanophotonics, Faculty of Science and Technology, MESA+ Institute, University of Twente, P.O. Box 217, 7500 AE Enschede, the Netherlands}
\affil[2]{Optical Technologies Group, Institute of Applied Physics, University of Münster, Corrensstraße 2, 48149 Münster, Germany}

\affil[*]{l.xia@utwente.nl}

\begin{abstract}
We investigate on-chip spectral control of supercontinuum generation, taking advantage of the additional spatial degree of freedom in strongly-coupled dual-core waveguides. Using numerical integration of the multi-mode generalized nonlinear Schrödinger equation, we show that, with proper waveguide cross-section design, selective excitation of supermodes can vary the dispersion to its extremes, i.e., all-normal or anomalous dispersion can be selected via phase shifting in a Mach-Zehnder input circuit. The resulting control allows to provide vastly different supercontinuum spectra with the same waveguide circuit.
\end{abstract}

\setboolean{displaycopyright}{false} % Do not include copyright or licensing information in submission.

\begin{document}

\maketitle

\section{Introduction}
%Lisi14June: I think it should start with two extreme types of SCG(with the corresponding applications) as early as possible but I have no idea how to arrange that at present*****************
Photonic waveguide circuits have strongly enhanced the control of gain and output spectra in on-chip nonlinear optical conversion. Examples are ultrahigh efficiencies in second-harmonic generation ~\cite{LoncarandFejerDOI10.1364/OPTICA.5.001438}, spectral shaping of four-wave mixing ~\cite{EppingDOI10.1364/OE.21.032123} or Kerr comb generation with integrated low-power diode pumping  ~\cite{GaetaDOI10.1038/s41586-018-0598-9}. Most challenging is controlling supercontinuum generation (SCG), where intense ultra-short optical input pulses are converted into broadband radiation in a single pass through a waveguide ~\cite{Jankowski2020}. A central ingredient to spectral control is based on imposing phase matching by proper control of the waveguide dispersion. In integrated photonic waveguides, this is achieved with lithographic dispersion engineering, leveraged by a high index contrast between core and cladding materials. For instance, in silicon nitride waveguides, anomalous dispersion can be induced such as to provide broadband coherent radiation spanning more than an octave ~\cite{BacheDOI10.1364/OL.41.002719,Epping2015,PorcelDOI10.1364/OE.25.001542}. Alternatively, the cross-section can be fabricated for all-normal dispersion, such as to render the broadened bandwidth suitable for temporal compression ~\cite{Torres-CompanyDOI10.1364/OE.450987}.    

%For instance, in silicon nitride anomalous dispersion can be induced across wide spectral ranges to open up broadband supercontinuum generation (SCG) for pre-selected pump wavelengths ~\cite{Epping2015,PorcelDOI10.1364/OE.25.001542}. %!Most challenging is controlling supercontinuum generation, where intense ultra-short optical input pulses are converted into a single pass through a waveguide, such as to provide broadband coherent radiation spanning more than an octave  !

To enable control of supercontinuum spectra also after lithographic fabrication, essentially four types of approaches have been investigated so far. Straightforward spectral shaping can be achieved with subsequent filtering such as with prisms ~\cite{PeterCimalla2009} or waveguide circuits ~\cite{Neshev2007},
however, this reduces the pump-to-output conversion efficiency. A second, more complex set of approaches exerts control of SCG via manipulating the pump radiation. Examples are tuning the pump wavelength ~\cite{NiklasM2021,Song2022DOI10.1038/s41598-022-22463-y}, pumping at several wavelengths simultaneously ~\cite{Brinkmann2016,Champert2004}, or varying the pump pulse duration ~\cite{M.Andreana2012}, pulse shape ~\cite{MorandottiandKuesDOI10.1038/s41467-018-07141-w}, and the pulse energy ~\cite{Genty2009,YuxingTang2016,FWang2019,LeiZhang2013}. %I removed the variation of the pump polarization ~\cite{NiklasandMaximilian2021} from here because this belongs to "multimode" SCG, as it needs at least two modes to vary the polarization.%
These methods provide shaping of SCG within certain ranges, however, alteration of the pump radiation is generally unwanted, as a change of pulse energy, spectrum and pulse shape is technologically involved. A third way of spectral controlling supercontinuum is to tune the material dispersion. A straighforward example is tuning the temperature ~\cite{Wang2018} such as demonstrated in liquid-core fibers. However, in solid-state materials of widely transparent waveguides, the effect is rather weak because thermo-optic coefficients are small, in the order of 10$\mathbf{^{-5}}$ (RIU/$^{\circ}$K) for Si$\mathbf{_{3}}$N$\mathbf{_{4}}$ and SiO$\mathbf{_{2}}$ ~\cite{Arbabi2013}. An optical control of dispersion, on ultrashort time scales, has been shown using the intensity dependence of the refractive index ~\cite{Melchert2020All-opticalSwitching}. However, such Kerr effects are also small and efforts are high, as a second, intense and independently controllable laser is to be injected.  

More recently, spectral control of SCG has begun to include the most powerful approach based on spatially multimode waveguides; here the additional degrees of freedom in tailoring optical fields are expected to facilitate a new class of complex nonlinear optical devices ~\cite{WrightLogan2022}.  %The unique advantage of the multimode waveguide systems strong modal dispersion~\cite{Li2018} .
Corresponding experiments are, however, constrained so far to bulk-fiber approaches controlled with free-space optics ~\cite{WrightPhysicsofhighlymultimode2022}. For instance, a wide range of largely different supercontinuum spectra has been demonstrated in a multimode fiber by varying the translation-stage alignment for the input beam ~\cite{Hu2003TunableCore}. However, due to the bulk approach associated with fiber input coupling, exciting particular spatial modes required adjustments of free-beam input focusing and direction. Drawbacks are then that selectivity in transverse excitation is limited by free-space diffraction, while oblique beam injection is rather alignment sensitive, introducing restrictions on long term stability.

 \begin{figure}[t]
\centering
\includegraphics[width=\linewidth]{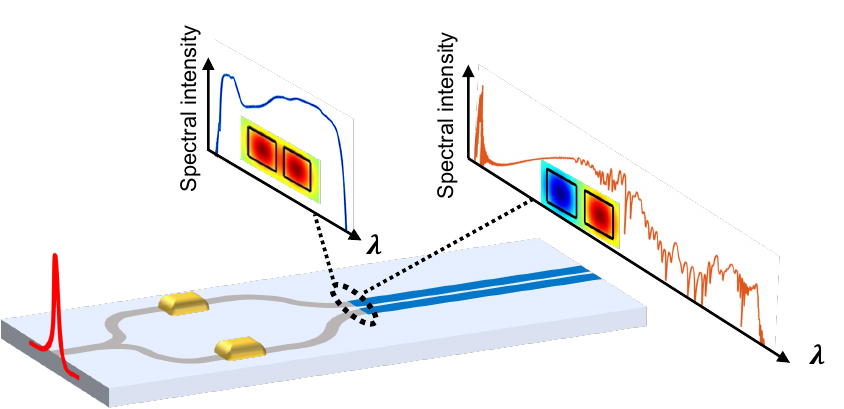}
\caption{Configuration for control of SCG in the Si$\mathbf{_{3}}$N$\mathbf{_{4}}$ dual-core waveguide (colored in blue) integrated with a Mach-Zehnder input circuit (grey). The phase shifters (yellow) integrated on the two arms of the input circuit can select in which transverse supermode supercontinuum is generated. }
\label{fig:device}
\end{figure}

Here we propose a novel path toward strong control of SCG suitable for a chip-integrated approach. As depicted in Fig. \ref{fig:device}, a Mach-Zehnder waveguide input circuit (colored in grey) in front of a strongly coupled dual-core waveguide section (blue) selects in which transverse supermode the supercontinuum is generated. We present numerical investigations based on the multi-mode generalized nonlinear Schrödinger equation to show that configuring the input circuit and choosing a proper design for the dual-core waveguide cross-section allows to select between two extremes of dispersion in supercontinuum generation. Using phase shifters (yellow) for addressing the symmetric supermode (SM) designed for all-normal dispersion generates spetcrally smooth and condensed supercontinuum output (see blue spectrum). Addressing the anti-symmetric supermode (AM)  designed for strongly anomalous dispersion generates a much wider spectrum, though with a more complex structure (see orange spectrum). We observe robustness of the approach as deviations from desired phase adjustments do not introduce additional complexities through intermodal nonlinear coupling. Benefiting from the on-chip spatial mode selection with a Mach-Zehnder type of input circuit, the described input coupling of pump pulses is solely to a single spatial mode at the chip entrance facet. However, following the input coupling, different transverse modes can be addressed via on-chip phase shifting, enabling the spectral control of SCG.

%Essentially, the inherent path length stability of integrated photonic circuits is used to render spectral control of supercontinuum generation more stable.

\section{Input circuit and dual-core waveguide}

\begin{figure}[t]
\includegraphics[width=0.9\linewidth]{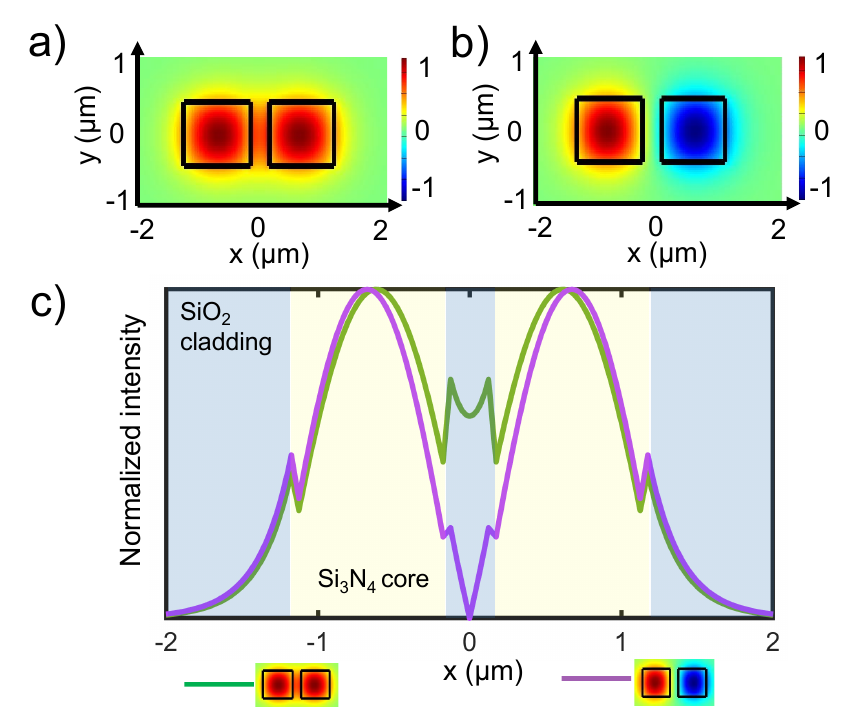}
\centering
\caption{a) Modeled E$\mathbf{_{x}}$ field distribution of the SM at 1554 nm. b) Modeled E$\mathbf{_{x}}$ field distribution of the AM at 1554 nm. c) Normalized field strength of the SM (green solid curve) and the AM (violet solid curve) at 1554 nm. The blue-shaded area and the yellow-shaded area represent the SiO$\mathbf{_{2}}$ cladding and the Si$\mathbf{_{3}}$N$\mathbf{_{4}}$ core, respectively.}
\label{fig:field distribution}
\end{figure}

For definiteness and to provide an example that may be implemented in an experiment covering wide spectral ranges, we consider the silicon nitride waveguide platform with Si$\mathbf{_{3}}$N$\mathbf{_{4}}$ cores embedded in a SiO$\mathbf{_{2}}$ cladding. The first part of the circuit consists of an adiabatically tapered single-spatial mode waveguide, allowing an ultrashort input pulse to be injected with high efficiency via free-space, fiber-to-chip or chip-to-chip coupling. To avoid any significant premature supercontinuum generation, the light intensity is kept sufficiently low with a bigger mode field diameter (in the order of 10 µm as in standard single mode fibers) via a thin core cross-section. Next the input pulse is divided into two approximately equal fractions that travel independently through the two arms of the input circuit. To adjust the relative phase of light in the two arms, and to compensate for potential fabrication errors in the path length difference, for instance thermoelectric heaters can be considered. Using two heaters in parallel has the advantage that at most half of the heating power is needed compared to a single heater. Due to the short length of the phase shifters, the wavelength dependence of the induced phase difference between the two arms can safely be neglected across the bandwidth of the input pulses. For instance, we calculate that the relative phase difference is less than 0.5\% across an 8 nm-bandwidth at 1554 nm wavelength. After phase shift adjustment, the two arms are adiabatically brought in proximity and tapered to tight-guiding thick-cores, thereby increasing the on-axis intensity. As the two tapered waveguides approach each other, the two formerly individual mode fields hybridize into transverse supermodes of the dual-core waveguide~\cite{Burns1988}. We note that letting two equally dimensioned cores approach adiabatically as in standard directional couplers ~\cite{Mrejen2015ADC, Ramadan1998ADC} ensures that, in the dual-core section, only two fundamental supermodes become noticeably excited, which are the fundamental symmetric supermode (SM) and the anti-symmetric supermode (AM) as depicted in Fig. \ref{fig:device}. Describing the optical fields in a dual supermode basis~\cite{Burns1988} serves for correctly including the strong linear coupling and also eases numerical modelling.

\begin{figure}[t]
\includegraphics[width=0.9\linewidth]{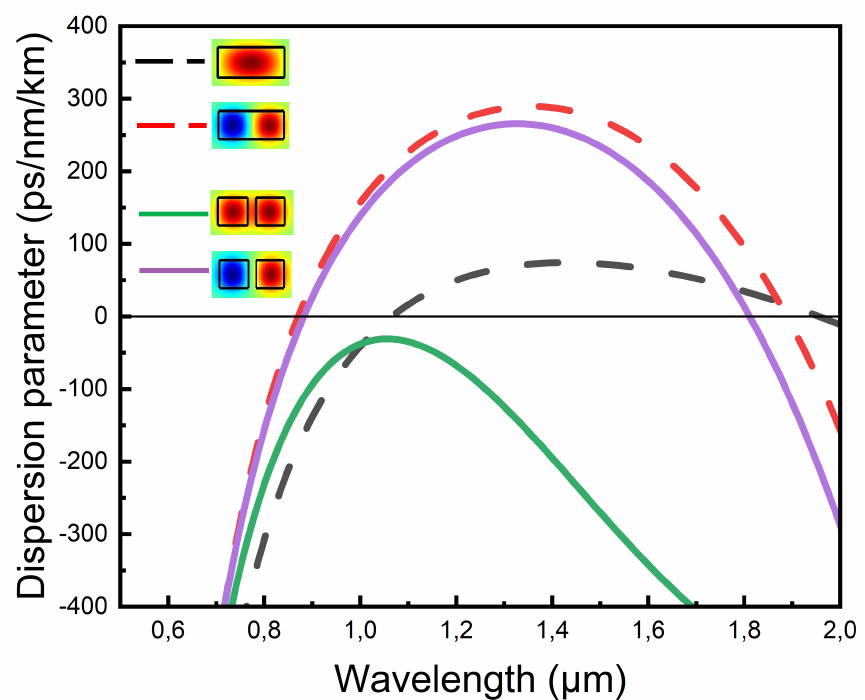}
\centering
\caption{Dispersion parameter profiles. For the dual-core waveguide, the SM (green solid curve) possesses all-normal dispersion and the AM (violet solid curve) possesses anomalous dispersion.  For a single-core waveguide (2.3 µm by 800 nm), the TE$\mathbf{_{00}}$ mode (black dashed curve) and the TE$\mathbf{_{10}}$ mode (red dashed curve) both posses anomalous dispersion.}
\label{fig:dispersion}
\end{figure}

To calculate the transverse field distributions of the two fundamental supermodes and the corresponding dispersion parameter profiles for the investigated core cross-sections, we use a 2D mode solver (Ansys Lumerical FDTD). To provide fully developed supercontinuum generation on chip-sized (sub-centimeter) interaction lengths, we start with a typical single-core cross-section (1 µm by 800 nm). In order to induce hybridization into two distinct supermodes, two such identical cores are placed in close proximity. Choosing a small horizontal gap of 300 nm between the cores is found to maximize the difference in the mode field distributions and dispersion of the supermodes. Figures \ref{fig:field distribution} a) and \ref{fig:field distribution} b) show the field distributions for the SM and the AM supermodes, respectively. The color code indicates the sign and strength of the electric field distributions.
% In the dual-core waveguide adopted in the work, two identical Si$\mathbf{_{3}}$N$\mathbf{_{4}}$ waveguide cores embedded in SiO$\mathbf{_{2}}$ are chosen as 1 µm (width) by 800 nm (height) and separated by a small horizontal gap of 300 nm. With such cross-sectional geometry of the two identical waveguide cores and their spacing, strongly different dispersions for the two supermodes in the pump wavelength range are imposed. The field distribution of the SM and the AM and their dispersion are found with a 2D mode solver. 
The calculated effective mode area at 1554 nm wavelength for the SM is 1.91 µm$\mathbf{^{2}}$ with 80\% of the light confined in the Si$\mathbf{_{3}}$N$\mathbf{_{4}}$ core. And for the AM, the effective mode area is 1.74 µm$\mathbf{^{2}}$ with 84\% of the light confined in the Si$\mathbf{_{3}}$N$\mathbf{_{4}}$ core. Figure \ref{fig:field distribution} c) provides a closer inspection of the light distribution in the materials. 
%As observed, the SM has slightly more light propagating in the Si$\mathbf{_{3}}$N$\mathbf{_{4}}$  cores than the AM, moreover, it has much more light propagating in the SiO$\mathbf{_{2}}$ cladding in between two Si$\mathbf{_{3}}$N$\mathbf{_{4}}$ cores than the AM, as indicated in the red dashed box in Fig. \ref{fig:field distribution} c). 
As a result of the difference in distribution, the two supermodes are expected to exhibit significantly different dispersion. The corresponding second-order dispersion parameters are displayed in Fig. \ref{fig:dispersion}. As observed, the SM and the AM in the dual-core waveguide offer opposite signs of the dispersion with the SM and the AM providing all-normal dispersion (green solid curve) and strong anomalous dispersion (violet solid curve), respectively. To highlight the merits of the dual-core vs a single-core approach, the dispersion of the TE$\mathbf{_{00}}$ and the TE$\mathbf{_{10}}$ modes for a single-core of the same overall dimensions (2.3 µm by 800 nm) are also given in Fig. \ref{fig:dispersion}. It can be seen that these modes are symmetric and anti-symmetric as well and thus possess different dispersion profiles. However, the sign of the dispersion is the same within much of the displayed wavelength range, such that the range of phase-shift induced change in dispersion is much smaller, and also the wavelength range where the sign can be altered is much smaller. 
%It is the same case for the TE$\mathbf{_{00}}$ and the TE$\mathbf{_{10}}$ modes in a single-core waveguide (2 µm by 800 nm) which provides the same amount of Si$\mathbf{_{3}}$N$\mathbf{_{4}}$ as the dual-core waveguide.
% Lisi_7June: should I add the following: such that firstly the SCG control range is relatively narrow and secondly there is high chance of nonlinear coupling between the two modes due to small. 
In contrast, the two supermodes in the dual-core waveguide offer opposite signs of dispersion parameters over a wide range such that vastly different supercontinuum spectra are expected.
%Lisi7June: should I add the following: the dispersion profiles of the SM and the AM for a single-core waveguide (2 µm by 800 nm) with the same volume of Si$\mathbf{_{3}}$N$\mathbf{_{4}}$ material as the dual-core waveguide have also been considered. However, similar to the dispersion profiles for a single-core waveguide (2.3 µm by 800 nm), the SM and the AM basically provide the same type of dispersion profiles.***************

\begin{figure}[t]
\includegraphics[width=\linewidth]{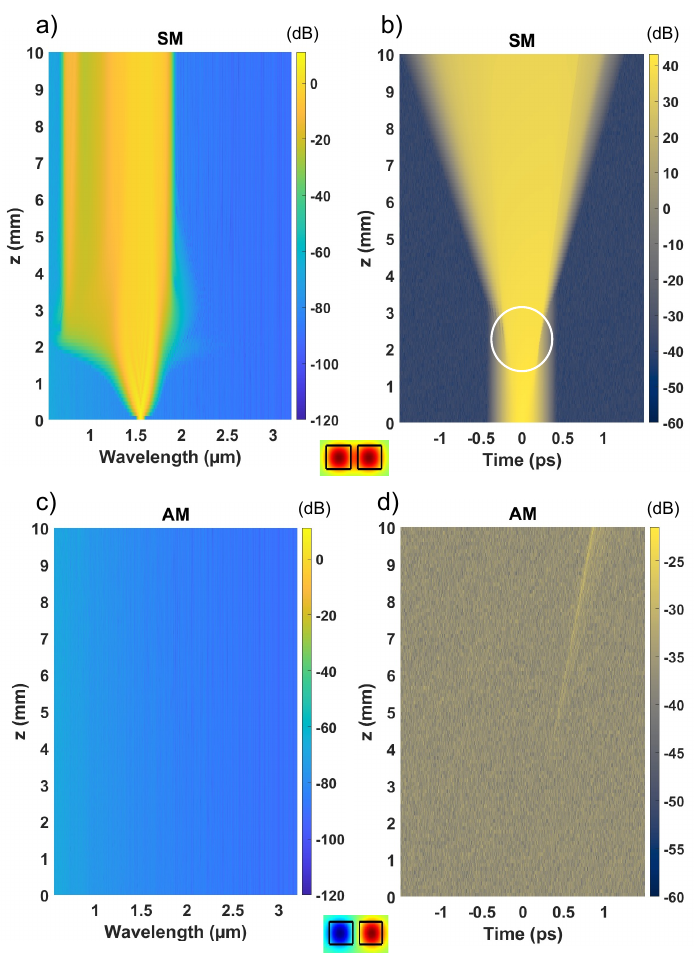}
\caption{a) Spectral evolution of SCG when only the SM is pumped. b) Temporal evolution of the SM. The temporal frame is chosen to move with the group velocity of the pump pulse in the SM. The white circle indicates where optical wave breaking occurs. The according spectral (c) and temporal evolution (d) in the non-pumped mode (AM) does not display any noticeable radiation.}
\label{fig:SCall-normal dispersion}
\end{figure}

\section{supercontinuum modelling}

To model SCG in the SM and the AM we use the multi-mode generalized nonlinear Schrödinger equation (MM-GNLSE) ~\cite{Poletti2008DescriptionFibers}  which is prepared to include higher-order dispersion, self-steepening, Raman scattering and nonlinear coupling of the modes. The latter, via intermodal effects such as cross-phase modulation (XPM) and four-wave mixing (FWM), is accounted for by the intermodal nonlinear coupling constants calculated from mode field overlap integrals ~\cite{Poletti2008DescriptionFibers, NiklasandMaximilian2021Doi10.1002/lpor.202100125}. Spontaneous generation of light is implemented as white noise fields set to the strength of the vacuum field in both modes. Integration of the equation is carried out assuming pumping with an Er-doped fiber laser system using as example a 1554-nm pump wavelength with a Gaussian pump pulse of 165 fs duration (corresponding to 8 nm bandwidth). The integration is numerically implemented over a propagation length of 10 mm. Using pulse energies in the range of 1 nJ, such length is sufficient to bring the spectral broadening to its saturation, meaning that the output spectra do not significantly broaden any more with further increasing the input power. 

\begin{figure}[t]
\includegraphics[width=\linewidth]{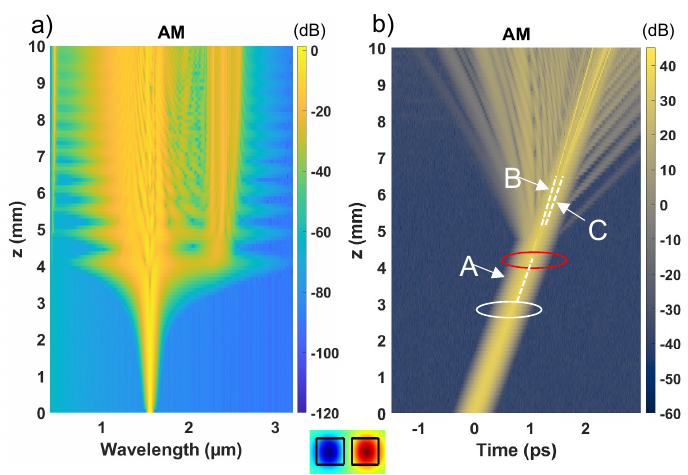}
\caption{a) Spectral evolution of SCG in the AM when only that mode is pumped. b) Temporal evolution of the AM. The temporal frame is chosen to move with the group velocity of the pump pulse. The white circle indicates where soliton A emerges; the red circle indicates where soliton fission occurs, splitting soliton A into fundamental solitons B and C.}
\label{fig:SCanomalous dispersion}
\end{figure}

\section{Results}

First, we consider pumping only the symmetric supermode (SM) that provides all-normal dispersion. Experimentally, this mode of excitation would be selected by setting the heater phase difference, $\Delta$$\phi$, between the two Mach-Zehnder arms in the input circuit equal to zero. The effect of deviations from such ideal setting is discussed further below. Figures \ref{fig:SCall-normal dispersion} a) and \ref{fig:SCall-normal dispersion} b) show, respectively, the spectral and temporal evolution of SCG in the SM as a function of the propagation length. A pulse energy of 2 nJ (11 kW peak power) is chosen to let the spectral evolution become well-completed within the 10 mm propagation length. Figures \ref{fig:SCall-normal dispersion} c) and \ref{fig:SCall-normal dispersion} d) show, for comparison, the evolution of the non-pumped mode (AM). It can be observed that a supercontinuum is generated in the pumped mode, while the non-pumped mode contains only spurious radiation (the spectrally integrated pulse energy fraction amounts to 1.6 x 10$\mathbf{^{-6}}$). In the pumped mode, the following essential dynamical features for the SCG in the SM can be identified: i) at the beginning of propagation, self-phase modulation (SPM) is the main contributor to the spectral broadening; ii) at a distance of around 2 mm, optical wave breaking occurs at the leading pulse edge and gives rise to a new spectral band extending toward short wavelengths down to 600 nm; iii) after around 4 mm propagation, the spectrum becomes continuously smoother. A -30-dB spectral bandwidth of about 1300 nm is achieved with a flat and smooth shape.

%For the first case of pumping only the SM, the spectral broadening reaches its saturation with a pulse energy of 2 nJ, corresponding to a peak power of 11 kW. It is found that a supercontinuum is generated in the pumped mode, while the non-pumped mode contains only spurious radiation (a fraction of 1.6 x 10$\mathbf{^{-6}}$ is calculated for the AM after the 10 mm interaction length). The spectral evolution and temporal evolution profiles of the supercontinuum in the SM as a function of propagation distance is shown in Figure \ref{fig:SCall-normal dispersion} a) and Figure \ref{fig:SCall-normal dispersion} b), respectively. The essential dynamical features can be identified: 

For the other case where pump radiation is injected only into the anti-symmentric mode (AM), $\Delta$$\phi$ in the input circuit is set equal to $\pi$. The spectral and temporal evolution profiles of the supercontinuum in the pumped mode are shown in Figs. \ref{fig:SCanomalous dispersion} a) and b), respectively. For pumping the AM, we find that lower pulse energies are sufficient for full spectral evolution within 10 mm, shown an example of 0.4 nJ and 2.3 kW peak power. The reason is the rather different dynamics: i) at the beginning, SPM is the main contributor to the broadening dynamics; ii) at around 3 mm a higher-order soliton emerges (labeled A, calculated soliton-order of 2) ; iii) with further propagation, soliton fission appears at a distance of around 5 mm such that the higher-order soliton splits into fundamental solitons (B and C), while dispersive waves lead to extensively broadening of the spectrum. As observed, the spectrum is much wider (more than 2500 nm), as expected due to the stronger confinement of the mode in the waveguide cores and due to the anomalous dispersion offered by the AM. We note that the tilt of the trajectories of the pulses in Fig. 4 b) is due to the different group velocity of the pulse in the AM compared to the group velocity in the SM which is chosen for the co-moving time frame. In Fig. \ref{fig:SCanomalous dispersion} we have displayed only the evolution in the pumped mode because we find that the energy in the non-pumped mode (SM) is negligible again (a fraction of 1.5 x 10$\mathbf{^{-5}}$).

\section{Discussion}
To explore the robustness of the scheme, the numerical modelling was extended to the case of simultaneous pumping of various linear superpositions of the AM and the SM. Such situations represent imperfect phase adjustment in the Mach-Zehnder waveguide input circuit, i.e., deviations from setting $\Delta$$\phi$ equal to zero or $\pi$, as might arise due to experimental imperfections. With imperfect phase settings, SCG in all-normal and anomalous dispersion can occur simultaneously and the two SCG processes may influence each other in a complex manner that is difficult to control, mainly via intermodal FWM and intermodal dispersive wave generation ~\cite{NiklasandMaximilian2021Doi10.1002/lpor.202100125}. The biggest possible deviation from pumping a single supermode occurs when $\Delta$$\phi$ is maximally different from 0 and $\pi$, i.e., $\Delta$$\phi$ = $\pi$/2. In this case, the SM and the AM equally share the pump radiation. To illustrate such an example we chose a total pulse energy of 0.8 nJ to synchronously excite both supermodes with each 0.4 nJ, while all other parameters are maintained.
%However, it is found that even with imperfect phase shifts with laser pulses (1554-nm pump wavelength with a pump pulse duration of 165 fs) the output remains a simple superposition of the individual spectra from the SM and the AM and can be distinguished. 
\begin{figure}[t]
\includegraphics[width=\linewidth]{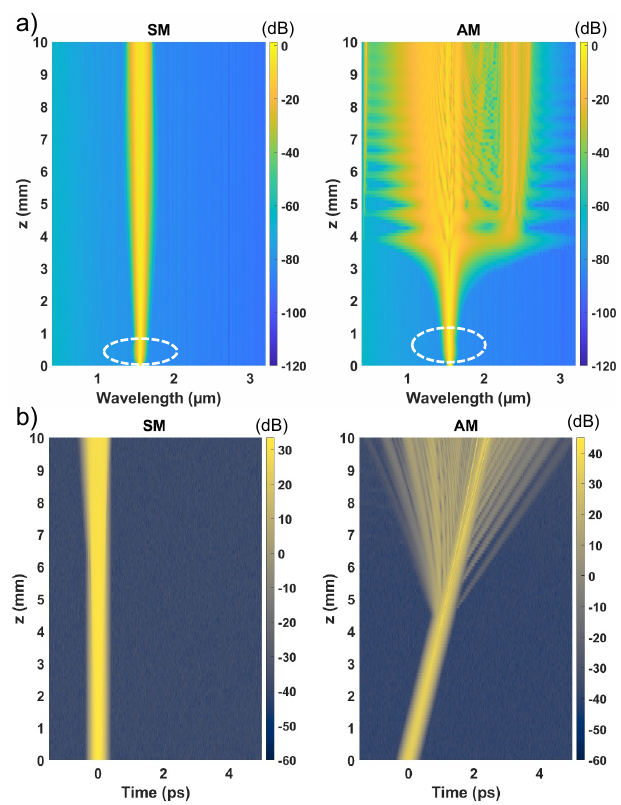}
\caption{ a) Evolution of supercontinuum spectra in the SM and the AM when pumped with equal pulse energies (phase difference setting $\Delta$$\phi$ = $\pi$/2). The white circles indicate where the pulses suffer a little distortion. b) Temporal evolution of SCG the SM and the AM. The temporal frame is chosen to move with the group velocity of the pump pulse in the SM.}
\label{fig:the SM and the AM equally sharing pump radiation}
\end{figure}

\begin{figure}[ht!]
\includegraphics[width=0.8\linewidth]{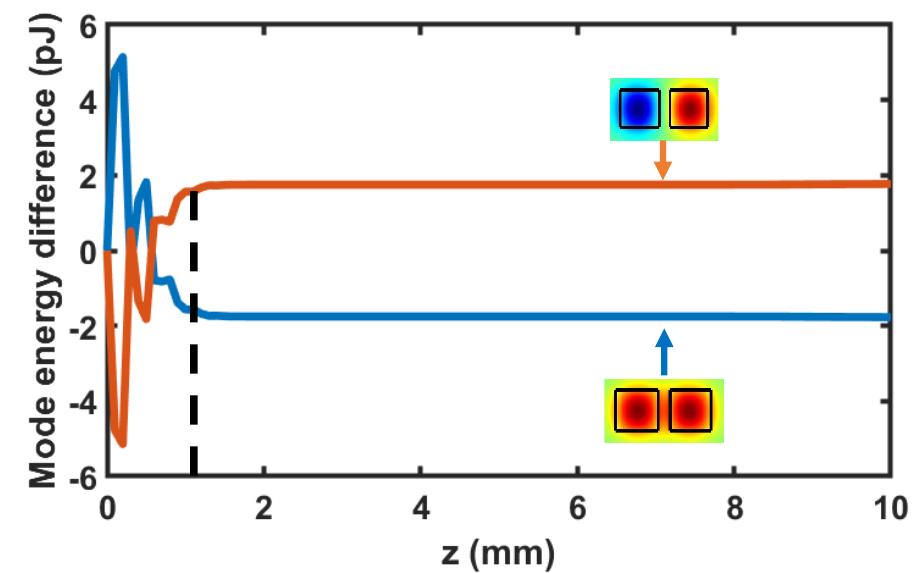}
\centering
\caption{Calculated change of mode energy in the SM and the AM vs propagation length for equal initial pulse energies (phase difference setting $\Delta$$\phi$ = $\pi$/2, initially 0.4 nJ pump energy in each mode). The black dashed line indicates the termination of energy exchange between the SM and the AM.}
\label{fig:mode energy differences}
\end{figure}

The spectral and temporal evolution is given in Fig. \ref{fig:the SM and the AM equally sharing pump radiation}. It can be seen that there is supercontinuum generation in both supermodes, and that the separate spectra and dynamics are similar to what is observed with pumping only a single supermode at a time. To analyze dual-mode pumping with regard to energy exchange by intermodal nonlinear interaction, Fig. \ref{fig:mode energy differences} plots on a finer scale how much the pulse energy of each mode changes with propagation. It can be seen that there is some energy exchange. However, the exchange is rather weak, at the pJ-level, which is in the order of 1\% of the pump pulse energy (0.4 nJ per mode). A further feature is that the exchange is only transient, limited to the initial dynamics in the first millimeter. In the spectral domain (white circles in Fig. \ref{fig:the SM and the AM equally sharing pump radiation} a) the intermodal interaction is barely visible and in the temporal domain (b) it appears essentially absent. Beyond about 1 mm-propagation length, there is no longer an interaction between the supermodes.

\begin{figure}[t]
\includegraphics[width=\linewidth]{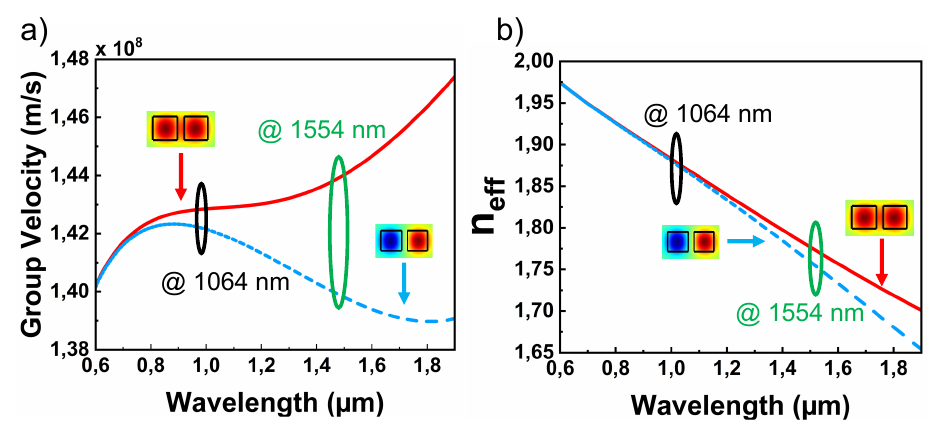}
\caption{a) Group velocity profiles and b) effective indices as a function of wavelength for the SM and the AM. Two standard pump wavelengths are indicated.}
\label{fig:Phase and group velocities}
\end{figure}

Such decoupling from nonlinear interaction can best be explained via the different temporal evolution profiles given in Fig. \ref{fig:the SM and the AM equally sharing pump radiation} b) where different tilts of trajectories indicate different group velocities. For further clarification, the absolute group velocities of the SM and the AM are plotted in Fig. \ref{fig:Phase and group velocities} a) as extracted from the dispersion of the effective indices in b). At the considered pump-wavelength of 1554 nm it can be seen that there is a significant velocity mismatch amounting to about 3\% of the absolute light velocity, with pulses in the AM lagging behind. As a result, the initial temporal overlap required for nonlinear intermodal effects quickly vanishes with propagation. Therefore, even with the maximum deviation from single supermode pumping assumed here, the total output remains to good approximation a simple superposition of the individual supercontinuum spectra. The processes don't mingle in time, or space nor in the spectral domain and therefore remain well distinguishable. To quantify the degree of residual nonlinear coupling for cases that would be more typical when experimentally aiming on single-supermode pumping, we also investigated SCG when the SM or the AM is dominantly excited with only some small fraction exciting the other mode. In the calculations, we find that intermodal nonlinear interaction becomes negligible. For instance with about 95\% of the pump radiation in one of the supermodes the energy exchange is typically less than 3 x 10$\mathbf{^{-4}}$ of the total energy. We note, however, that the calculated nonlinear decoupling of SCG in two supermodes becomes progressively weaker when shifting to a shorter pump wavelength (for instance 1 µm or closer toward the visible) with the given dual-core cross section. This can be indeed be expected based on Fig. \ref{fig:Phase and group velocities} where the group velocities and phase velocities of the SM and AM approach each other toward shorter wavelength.

\section{Conclusion}
To conclude, we propose an approach for controlling integrated waveguide supercontinuum generation via supermode selection solely with on-chip phase tuning. The approach is based on transverse spatial mode dynamics in strongly coupled dual-core waveguides. Thereby, simple phase tuning can be used to vary the dispersion to its extremes, i.e., to select either all-normal (all positive) dispersion, or strong and wide anomalous (negative) dispersion, and thus generate vastly different output spectra. Exploiting strong linear coupling in a dual-core waveguide vs. a single-core transverse mode approach maximizes the difference in mode dispersion. This also minimizes the nonlinear coupling between the symmetric and anti-symmetric supermodes, which is found to be negligible for the considered case of pumping with ultrashort pulses from Er-lasers (1554 nm) when optimizing the dual core waveguide cross section. Maximizing the difference in group velocities also provides a significant robustness of supercontinuum generation vs. imperfect phase tuning in supermode selection. Controlling supercontinuum over wide ranges, via simple on-chip phase shifting to access either anomalous dispersion or all-normal dispersion may bring increased flexibility to integrated photonics applications involving configuring of broadband light ~\cite{Song2022DOI10.1038/s41598-022-22463-y}. Potentially the approach can also be extended to systems with multicore waveguide-arrays by configuring Mach-Zehnder tree input circuits with multiple phase shifters.

\section*{Funding} This work is partially funded by the Dutch Science Organization, NWO, within the Synoptics program, P17-1.

\section*{Acknowledgments} The authors would like to thank C. Roeloffzen for useful discussion and inspiration regarding waveguide design for the input circuit.

\section*{Disclosures} The authors declare no conflicts of interest.

% Bibliography
\bibliography{Citation}

% Full bibliography added automatically for Optics Letters submissions; the following line will simply be ignored if submitting to other journals.
% Note that this extra page will not count against page length

%\bibliographyfullrefs{sample}

%Manual citation list
%\begin{thebibliography}{1}
%\bibitem{Zhang:14}
%Y.~Zhang, S.~Qiao, L.~Sun, Q.~W. Shi, W.~Huang, %L.~Li, and Z.~Yang,
 % \enquote{Photoinduced active terahertz metamaterials with nanostructured
  %vanadium dioxide film deposited by sol-gel method,} Opt. Express \textbf{22},
  %11070--11078 (2014).
%\end{thebibliography}

\end{document}